# Room-temperature ferromagnetism at an oxide/nitride interface

**Authors**

Qiao Jin[1,2,†], Zhiwen Wang[3,†], Qinghua Zhang[1,†], Yonghong Yu[1,2], Shan Lin[1,2], Shengru Chen[1,2], Mingqun Qi[1], He Bai[4], Qian Li[5], Le Wang[6], Xinmao Yin[7], Chi Sin Tang[8], Andrew T. S. Wee[9], Fanqi Meng[1], Jiali Zhao[1], Jia-Ou Wang[10], Haizhong Guo[11], Chen Ge[1], Can Wang[1,2,12], Wensheng Yan[5], Tao Zhu[1,4,12], Lin Gu[1,2,12], Scott A. Chambers[6], Sujit Das[13], Gang-Qin Liu[1,2,12], Shanmin Wang[14], Kui-juan Jin[1,2,12,*], Hongxin Yang[3,*], and Er-Jia Guo[1,2,12,*]

**Affiliations**

[1] Beijing National Laboratory for Condensed Matter Physics and Institute of Physics, Chinese Academy of Sciences, Beijing 100190, China

[2] University of Chinese Academy of Sciences, Beijing 100049, China

[3] Ningbo Institute of Materials Technology & Engineering, Chinese Academy of Sciences, Ningbo 315201, China

[4] Spallation Neutron Source Science Center, Dongguan 523803, China

[5] National Synchrotron Radiation Laboratory, University of Science and Technology of China, Hefei 230029, China

[6] Physical and Computational Sciences Directorate, Pacific Northwest National Laboratory, Richland, WA 99354, USA

[7] Physics Department, Shanghai Key Laboratory of High Temperature Superconductors, Shanghai University, Shanghai 200444, China

[8] Institute of Materials Research and Engineering, A*STAR (Agency for Science, Technology and Research), 2 Fusionopolis Way, Singapore, 138634 Singapore

[9] Department of Physics, National University of Singapore, 2 Science Drive 3, Singapore 117551, Singapore

[10] Institute of High Energy Physics, Chinese Academy of Sciences, Beijing 100049, China

[11] Key Laboratory of Material Physics, Ministry of Education, School of Physics and Microelectronics, Zhengzhou University, Zhengzhou 450001, China

[12] Songshan Lake Materials Laboratory, Dongguan, Guangdong 523808, China

[13] Unité Mixte de Physique, CNRS, Thales, Université Paris-Saclay, Palaiseau, France

[14] Department of Physics, Southern University of Science and Technology, Shenzhen 518055, China

*Corresponding author. Emails: kjjin@iphy.ac.cn, hongxin.yang@nimte.ac.cn, and ejguo@iphy.ac.cn.

† These authors contributed equally to this work.

**Abstract:** Heterointerfaces have led to the discovery of novel electronic and magnetic states because of their strongly entangled electronic degrees of freedom. Single-phase chromium compounds always exhibit antiferromagnetism following the prediction of Goodenough-Kanamori rules. So far, exchange coupling between chromium ions via hetero-anions has not been explored and the associated quantum states is unknown. Here we report the successful epitaxial synthesis and characterizations of chromium oxide ($Cr_2O_3$)-chromium nitride (CrN) superlattices. Room-temperature ferromagnetic spin ordering is achieved at the interfaces between these two antiferromagnets, and the magnitude of the effect decays with increasing layer thickness. First-principles calculations indicate that robust ferromagnetic spin interaction between $Cr^{3+}$ ions via anion-hybridizations across the interface yields the lowest total energy. This work opens the door to fundamental understanding of the unexpected and exceptional properties of oxide-nitride interfaces and provides access to hidden phases at low-dimensional quantum heterostructures.

**Main text**

Artificially structured heterostructures are ideal platforms for observing emergent phenomena and monitoring dynamic properties evolution following external stimuli[1-3]. Of the new functionalities that have emerged, ferromagnetism generated from two non-ferromagnetic materials is fascinating from a fundamental perspective and is of considerable technological interest for various spintronics architectures[4,5]. Interactions between magnetically active ions can be estimated theoretically to determine whether heterostructures should exhibit ferromagnetism or antiferromagnetism. However, the discovery of ferromagnetism in superlattices (SL) comprising of two antiferromagnets[6-8], or an antiferromagnet and a paramagnet[9-12], is expected based on Goodenough-Kanamori rules[13,14]. Interfacial magnetism has been reported in all-oxide or metal/oxide heterostructures over the past two decades[6-12,15,16]. Several dominant exchange mechanisms have been proposed to explain this intriguing effect, including interfacial charge transfer, structural coupling, orbital reconstruction, and tailoring the exchange interaction via cation ordering[17-20]. However, emergent magnetic states at transition metal oxide/nitride interfaces have never been discussed previously in the literature. The challenge of fabricating stoichiometric nitride thin films lies in the precision of controlling stoichiometry and crystallinity in this class of emergent heterojunctions, as is also the case with perovskite oxide interfaces and epitaxial metal/oxide interfaces[21,22]. We have recently developed a novel method for depositing stoichiometric nitride films using plasma-assisted pulsed laser deposition[23,24], paving the way for the fabrication of artificially structured SLs comprised of transition metal nitrides and oxides.

Bulk CrN is a paramagnet with a cubic rock-salt structure at higher temperatures and transforms to a C-type antiferromagnet below the Néel temperature ($T_N$ ~280 K)[25]. $Cr_2O_3$ has a corundum structure, comprising a hexagonal close-packed array of oxygen ions, with 2/3 of the octahedral holes occupied by Cr. The antiferromagnetic order in $Cr_2O_3$ persists up to 307 K[26]. The Goodenough–Kanamori rules predict antiferromagnetic interaction between $Cr^{3+}$ cations in the pure materials[13,14]. The spins align antiparallel along the [110] orientation in CrN, whereas the antiferromagnetic spin texture in $Cr_2O_3$ is parallel to [0001]. The spin realignment or ordering at the planar interfaces of CrN and $Cr_2O_3$ has not been discussed in the literature and the underlying mechanism associated with magnetic interaction via dissimilar anions is unexplored so far. Here, we report the synthesis of all-Cr-based SLs consisting of alternating CrN and $Cr_2O_3$ layers. By employing multiple magnetic probes and theoretical calculations, we confirm that ferromagnetic coupling of the two materials occurs and persists up to ~325 K in short-period SLs. By controlling the SL periodicity, we determine that the averaged saturation magnetization decays as the layer thickness increases, revealing that the spin ordering is localized at the CrN-$Cr_2O_3$ interface. The functional coupling of nitrides and oxides will generate significant interest because of the emergent phenomena occurs at this type of quantum heterointerface.

Epitaxial $[(Cr_2O_3)_t/(CrN)_t]_{10}$ ($[CO_t/CN_t]_{10}$) SLs, where $t$ denotes the thicknesses of $Cr_2O_3$ and CrN layers in the unit of nm and ranges from 0.5 to 5, were grown on (0001)-oriented $Al_2O_3$ single-crystal substrates using plasma-assisted pulsed laser deposition (see Supplementary Experimental section). Synchrotron-based x-ray diffraction (sXRD) and topography measurements reveal that all SLs are of high crystalline quality, coherently strained, targeted layer thickness, and exhibit smooth surfaces (Figs. S1-S3). A scanning transmission electron microscopy (STEM) high-angle annular dark-field (HAADF) image of a $[CO_1/CN_1]_{10}$ SL is shown in Fig. 1a. The white spots in the image are columns of Cr atoms. The brighter spots originate from Cr atoms in CrN, whereas the dimmer spots are from Cr atoms in $Cr_2O_3$ because the density of Cr ions is higher in CrN than that in $Cr_2O_3$ along the incident beam direction. The CrN growth direction aligns with the [111] crystallographic orientation, whereas that of $Cr_2O_3$ is along [0001]; this orientation minimizes in-plane misfit strain. Along the growth direction, the distance between adjacent Cr-containing planes is ~2.40 and ~2.77 Å in CrN and $Cr_2O_3$, respectively. Therefore, the thicknesses of CrN and $Cr_2O_3$ layers are approximately 1 nm. To visualize the distribution of elements at the atomic scale, we conducted electron-energy-loss spectroscopy (EELS) mapping combined with STEM-HAADF over the entire SL (Fig. 1b). EELS mapping indicates the Cr ions are uniformly distributed (Fig. 1c), and N and O ions are well-separated in the SL (Figs. 1d-1f). Please note that the EELS profiles exhibit additional broadening due to beam delocalization resulting from anisotropic inelastic scattering as phonons are excited. This kind of broadening does not occur in HAADF because the more Rutherford-like scattering potential results in a higher extent of localization at the nuclei. Therefore, the separation of CrN and $Cr_2O_3$ layers can further be identified from the intensity contrast of Cr atoms due to the difference in atom density along the STEM projective direction (Fig. S4). We also analyzed the strain distribution within the SL (Fig. S5). Although the SL is fully in-plane coherently strained, the out-of-plane lattice constants alternatively change along the growth direction. Figs. 1g and 1h show that the spacings between the Cr planes in $Cr_2O_3$ and CrN layers are narrow distributions at distances of ~1.12 and 0.98 nm, respectively, with a standard deviation of 0.02 nm. These structural characterizations thus demonstrate that we successfully fabricated high-quality $Cr_2O_3$/CrN SLs with a high level of spatial uniformity.

We investigated the electronic states in $Cr_2O_3$/CrN SLs by conducting x-ray absorption spectroscopy (XAS) and x-ray photoelectron spectroscopy (XPS) measurements. The angular momentum coupled final states at the Cr $L_{3,2}$-edges (dashed lines in Fig. 1i) are characteristic of tri-valent Cr cations. XAS of our SL at the O $K$- and N $K$-edges exhibits the same lineshapes of those in the individual stoichiometric CrN and $Cr_2O_3$ single films, suggesting that no apparent O and N vacancies in SL. Angle-dependent XPS measurements for $[CO_{0.5}/CN_{0.5}]_{10}$ provide evidence of negligible charge transfer across the interfaces because the Cr $2p$ lineshape does not change with photoelectron probe depth (Fig. S6), indicating the same valence (3+) for Cr in all SL periods sampled at the various angles. Therefore, both techniques indicate that the magnetic and electronic properties changes in SLs are not driven by a change in Cr valence across the buried interfaces.

The impact of interfacial coupling between CrN and $Cr_2O_3$ layers is reflected in the field- and temperature-dependent magnetization ($M$) variations with SL periodicity. The SL shows strong magnetic anisotropy when the magnetic easy axis is along the in-plane direction (Fig. S7). The samples exhibit ferromagnetic behavior when the saturation magnetization ($M_S$) and coercive field ($H_C$) both reduce as the temperature increases. Fig. 2a shows the $M$-$H$ curves of $Cr_2O_3$/CrN SLs with different layer thicknesses at 10 K. $M$ is calculated by measuring moment and dividing by the volume of the entire sample. From these results, the SL with the thinnest layers has the largest $M_S$ but the smallest $H_C$ (Figs. 2b and 2c). The overall $M$ of $[CO_{0.5}/CN_{0.5}]_{10}$ is significantly larger than that of [CO$_1$/CN$_1$]$_{10}$ and the *M* value of is close to zero when *t* exceeds 1.5 nm. The overall magnetization appears to follow a $t^{-1.4}$ dependence. The strength of the interaction is strongly dependent on layer thickness, with the maximum effect occurring at a length scale of a few u.c. The *M-H* hysteresis loop at 300 K (inset of Fig. 2b) and *M-T* curves (Fig. S7) for [CO$_{0.5}$/CN$_{0.5}$]$_{10}$ reveal ferromagnetism up to ~325 K, which is larger than the $T_N$ of both parent materials. Similar behavior had been observed at the Cr$_2$O$_3$ nanoparticles with reduced dimensionality[27]. To exclude the influence of interfacial intermixing to the magnetic properties of SLs, we measured the field-dependent magnetization of a N doped Cr$_2$O$_3$, a O doped CrN, and a deliberately intermixed alloy film (Fig. S8). All of them exhibit nearly negligible magnetic moment and no hysteresis loop, suggesting that anion intermixing at the interfaces cannot explain the observed ferromagnetic response in the SLs.

To identify the intriguing origin of the magnetic properties of these SLs, we performed x-ray magnetic circular dichroism (XMCD) spectra measurements on [CO$_{0.5}$/CN$_{0.5}$]$_{10}$ at various temperatures (Fig. S9). Fig. 2c shows XAS at Cr $L_{3,2}$-edges under magnetic fields of ±0.7 T at 86 and 300 K. The calculated XMCD spectra at 86 K shows that the SL exhibits a large negative response at the $L_3$ edge and a positive response at the $L_2$ edge, indicating that the magnetic moments of Cr ions are oriented parallel to the applied fields (Fig. 2e). Finite XMCD signals at 300 K are observed, indicating persistent ferromagnetic ordering at room temperature. By applying sum rules to the Cr XMCD spectra, we can separate the spin (*S*) and orbital (*L*) contributions to the magnetization (Fig. S9). The *S* and *L* values for Cr in [CO$_{0.5}$/CN$_{0.5}$]$_{10}$ at 86 K are ~0.10 $\mu_B$/Cr and ~0.03 $\mu_B$/Cr, respectively. Thus, the total magnetic moment (*M*=2*S*+*L*) is ~0.23 $\mu_B$/Cr, agreeing with the magnetometry results. Additionally, the ferromagnetism in $Cr_2O_3$/CrN SLs was confirmed using noninvasive magnetometry of nitrogen-vacancy (NV) centers in nanodiamonds[28-30]. As shown in Fig. 3a, nanodiamonds with ensemble NV centers were dispersed on the surface of [$CO_{0.5}$/$CN_{0.5}$]$_{10}$ SL, and continuous-wave optically detected magnetic resonance (ODMR) spectra were measured under zero magnetic field to identify the stray fields of the sample. Fig. 3b shows typical ODMR spectra of a measured nanodiamond at various temperatures. The splitting between the two resonant peaks ($\omega_1$ and $\omega_2$), which is proportional to the local stray magnetic field (Zeeman effect), rapidly drops when warming from 89 to 295 K, indicating the remnant magnetization of the SL strongly depends on the temperature. As a comparison, we found that the ODMR splitting of nanodiamonds on a bare substrate ($Al_2O_3$) shows no temperature dependence (Fig. S10). The same NV-based magnetometry measurements were repeated randomly on at least twenty nanodiamonds. The results indicate that the net magnetic moment of [$CO_{0.5}$/$CN_{0.5}$]$_{10}$ at 89 K is statistically larger than that at 295 K. The trend in magnetization obtained from NV-center magnetometry correlates with $M$-$T$ curves obtained from macroscopic magnetization measurements (Fig. 3c), corroborating the origin of the ferromagnetic signals in the SLs.

To separate the magnetic contributions from the individual layers to the averaged $M$, polarized neutron reflectometry (PNR) was used to quantitatively determine the magnetization depth profile across the SLs. As shown in Fig. 3d, the specular reflectivity ($R$) of the SL was measured as a function of wave vector transfer $\vec{q}$. The neutron reflectivities from both spin states are normalized to the Fresnel reflection ($R_F$). $R^+$ and $R^-$ represent the reflectivities of neutrons with spin parallel and antiparallel to the applied field, respectively. Fig. 3d shows the calculated spin asymmetry (SA), ($R^+ - R^-$)/($R^+ + R^-$), of [$CO_{0.5}$/$CN_{0.5}$]$_{10}$ as a function of $\vec{q}$ at 10 K under a field of 1 T. The SA carries information on the depth variation of the net magnetization in the SL. By fitting the PNR data (Fig. S11) and SA to a model describing the chemical depth profile (Fig. S2)[31], we obtained the magnetic scattering length density (mSLD) corresponding to the net magnetization of $[CO_{0.5}/CN_{0.5}]_{10}$ as a function of film depth. Here we use a uniformly distributed magnetization within SLs because the PNR data collected over a limited $\vec{q}$ range are not of sufficient resolution to separate the magnetization contributions in the CrN and $Cr_2O_3$ layers with a thickness of 0.5 nm. In contrast, no satisfactory fit could be found if $M$ is constrained to zero, yielding firm evidence for the existence a net moment within SLs[32,33]. The thickness-averaged $M$ values at 10 K are $26.5 \pm 5.0$ emu/cm$^3$ for SLs. We measured the same sample at the BL-4A of spallation neutron source at ORNL after field cooling at 4.8 T and 10 K (Fig. S12). The averaged magnetization of $[CO_{0.5}/CN_{0.5}]_{10}$ is $31.8 \pm 7.0$ emu/cm$^3$, suggesting a saturation moment is achieved above 1 T. The $[CO_{2.5}/CN_{2.5}]_{10}$ SL exhibits a negligible magnetization under the same conditions, agreeing with the SQUID results that $M$ decreases with increasing layer thickness.

In the present work, we hypothesize that the strong orbital ordering between $Cr^{3+}$ ions that occurs through the nonmagnetic anions (N and O) triggers interfacial ferromagnetic ground states formation across the interfaces. The frustrated spin states in both $Cr_2O_3$ and CrN layers induce a large canting angle toward the interfaces, resulting in a formation of long-range spin ordering within the SLs. The strength of the interfacial coupling between two neighboring Cr ions via N and O anions directly controls the magnitude of the canting angle. For ultrathin layers, spins are readily confined to the plane of the interface, thus producing a large net moment preferentially aligned along the lateral interfaces. As the layer thickness increases, the interfacial coupling weakens due to the frustrated bulk layers. Eventually the antiferromagnetic ground states of the thicker $Cr_2O_3$ and CrN layers dominate the macroscopic magnetization, reducing the average moment to zero.

First-principles calculations reveal that the most stable magnetic configuration of the SL is the ferromagnetic state which exhibit the lowest formation energy, independent of interface structure and layer thickness (Figs. S13-S16 and Table S1) [34,35]. Figs. 4a-4c show the calculated band structures for SLs with $Cr_2O_3$ and CrN layer thicknesses ranging from 0.5 to 1 nm. For all SLs with $t \leq 1$, an indirect bandgap appears and the electronic structures for spin-up and spin-down configurations are degenerate, indicating that the intrinsic states for the SLs are both ferromagnetic and insulating. We find that the SL bandgap reduces as the layer thickness gradually increases from 0.5 to 1 nm (Fig. 4e). At the same time, the gap of the spin-up band closes while that of the spin-down band remains the same as the CrN layer thickness increases. Additionally, we find that the averaged magnetic moments of each Cr atom decrease with increasing thickness (Fig. 4d and Table S2). This result is consistent with the reduced net moment produced by spin canting along the in-plane direction. This fact supports that the interior portions of the thick layers exhibit frustrated magnetization. Thus, the driver behind the decreased magnetization is competition between exchange coupling at the interface and other terms in order to minimize the total energy. We note that the theoretical calculations remain some mismatches in the magnitude of measured magnetic moments at the interfaces. These can be further improved by precisely determining the exchange coupling strength across the interface and canting angle within the layers. In addition, the calculation results also agree with our transport measurements, indicating that all SLs exhibit insulating behavior with the conductance increasing with layer thickness (Fig. S17). As found in our earlier work,[23,24] the insulating nature of ultrathin CrN layers is attributed to the dimensionality-driven metal-to-insulator transition in the individual CrN single layers. We find that the thickness-dependent activated thermal energy quantitatively correlates with the evolution of the bandgap and follows a $1/t$ dependence. The calculations thus shed considerable light on our experimental observations and provide a theoretical underpinning to deepen our understanding of this class of heterointerfaces.

In summary, we have combined multiple magnetic probes and theoretical calculations to demonstrate emergent ferromagnetism in a class of prototypical epitaxial oxide/nitride superlattices, $[(Cr_2O_3)_t/(CrN)_t]_{10}$, that incorporate the same isovalent cation. We conclude that the canted spin texture induced by magnetic coupling at the various buried interfaces is the driver of this unique magnetic ground state. This fundamental study sheds light on interfacial coupling between functional oxides and nitrides, and further deepens our understanding of how to bridge these two important classes of materials with their similar lattice symmetries and unit cell sizes. Combining transition metal oxides and nitrides in this way will stimulate additional theoretical and experimental efforts that will trigger a deepened understanding of the role of such asymmetric interfaces. Finally, this work opens new avenues for potential applications involving a class of materials that heretofore have not been used in any spintronic technologies.

**Acknowledgments**

The authors thank Dr. G. Q. Yu and Dr. Z. H. Zhu at the Institute of Physics, Chinese Academy of Sciences, Prof. D. Z. Hou, Prof. L. F. Wang, and Prof. Z. Liao at the University of Science and Technology, Prof. Q. Li at Tsinghua University, and Prof. H. W. Guo at Fudan

University for valuable discussions. This work was supported by the National Key Basic Research Program of China (Grant Nos. 2019YFA0308500 and 2020YFA0309100), the National Natural Science Foundation of China (Grant No. 11974390, 52025025, 52072400), the Beijing Nova Program of Science and Technology (Grant No. Z191100001119112), Beijing Natural Science Foundation (Z190010 and 2202060), the Strategic Priority Research Program of Chinese Academy of Sciences (Grant No. XDB33030200), and the Guangdong-Hong Kong-Macao Joint Laboratory for Neutron Scattering Science and Technology. The PNR experiments were conducted at the Beamline MR of Chinese Spallation Neutron Scource, CAS, and partial PNR work were used resources at the Spallation Neutron Source, a DOE Office of Science User Facility operated by the ORNL. XAS experiments were conducted at the Beamline 4B9B of BSRF. XRD measurements were conducted at the beamline 14B1 and 02U2 of SSRF. XMCD measurements were conducted at NSRL in China and SSLS in Singapore. XPS measurements were performed at PNNL, a DOE Office of Science, Office of Basic Energy Sciences, the Division of Materials Sciences and Engineering under Award Nr. 10122.

**Figure Legends**

**Figure 1. Structural and electronic characterizations of $Cr_2O_3$/CrN SLs.** (a) A high-magnified STEM image from a region marked with orange dashed rectangle in (b), which is a cross-sectional STEM-HAADF image for a $[CO_1/CN_1]_{10}$ SL. The schematic of crystal structures across two adjacent interfaces is illustrated on the right side of STEM image. The bright and dark regions correspond to the CrN and $Cr_2O_3$ layers, respectively. The atomic distances between the Cr planes in CrN and $Cr_2O_3$ are ~2.40 and ~2.77 Å, respectively. The HAADF-STEM image was acquired along the $[\bar{1}100]$ zone axis. Integrated STEM-EELS intensity maps for the (c) Cr $L_{3,2}$-, (d) O $K$-, and (e) N $K$-edges, indicating a uniform distribution of elements across the interfaces. (f) EELS intensity profiles for O $K$- and N $K$-edges obtained from EELS line scans which were averaged across the entire image. Statistical distributions of Cr spacing planes between (g) the $Cr_2O_3$ and (h) CrN layers. (i) XAS for a $[CO_1/CN_1]_{10}$ SL, a CrN and a $Cr_2O_3$ single layer at N $K$-, O $K$-, and Cr $L_{3,2}$-edges.

**Figure 2. Magnetometry and XMCD measurements for $Cr_2O_3$/CrN SLs**. (a) In-plane *M-H* curves for symmetric $[CO_t/CN_t]_{10}$ SLs at 10 K. (b) $M_S$ and (c) $H_C$ as a function of layer thickness. Inset: *M-H* loops for $[CO_{0.5}/CN_{0.5}]_{10}$ SL at 300 K. (d) XAS at Cr $L$-edges for $[CO_{0.5}/CN_{0.5}]_{10}$ with in-plane fields of $\pm$ 0.7 T. (e) XMCD at 86 and 300 K.

**Figure 3. Nanoscale magnetic probing using NV spins in nanodiamond and PNR.** (a) Schematic of diamond NV-based magnetometry. (b) Zero-field optically detected magnetic resonance (ODMR) spectra of ensemble NV centers in a nanodiamond on the surface of $[CO_{0.5}/CN_{0.5}]_{10}$ SL at various temperatures. (c) *M-T* curves of $[CO_{0.5}/CN_{0.5}]_{10}$ at 1 kOe after zero-field cooling (open circles) and field cooling of 1 kOe (solid circles). The temperature

dependence of $2\gamma B$ is plotted as color circles for comparison. (d) Normalized neutron reflectivities (upper panel) from both spin-up and spin-down neutrons and spin asymmetry (lower panel) of $[CO_{0.5}/CN_{0.5}]_{10}$ SL as a function of $\vec{q}$. (e) Magnetization depth profile calculated from the fitted neutron magnetic scattering length density.

**Figure 4. Evolution of band structures with layer thickness.** (a)-(c) Band structures of the $Cr_2O_3$/CrN SLs with CO and CN layer thickness ranging from 0.5 to 1 nm, respectively. The black and red curves represent the spin-up and spin-down states, respectively. (d) Magnetic moments for the Cr, O, and N ions and (e) the bandgaps calculated for a series of $[CO_t/CN_t]$ SLs. The bandgap is defined by the difference between the eigenvalues at the bottom of conduction band and the top of valence band.

**Figures 1**

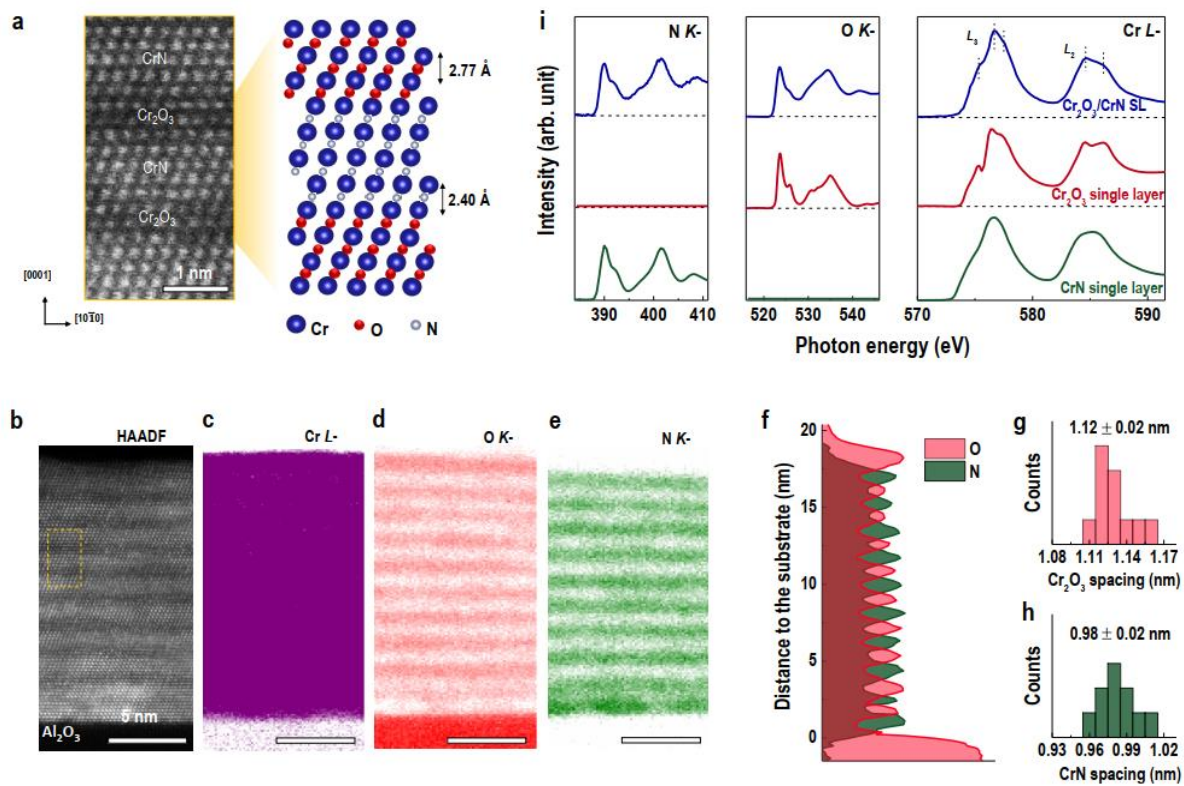

**Figure 2.**

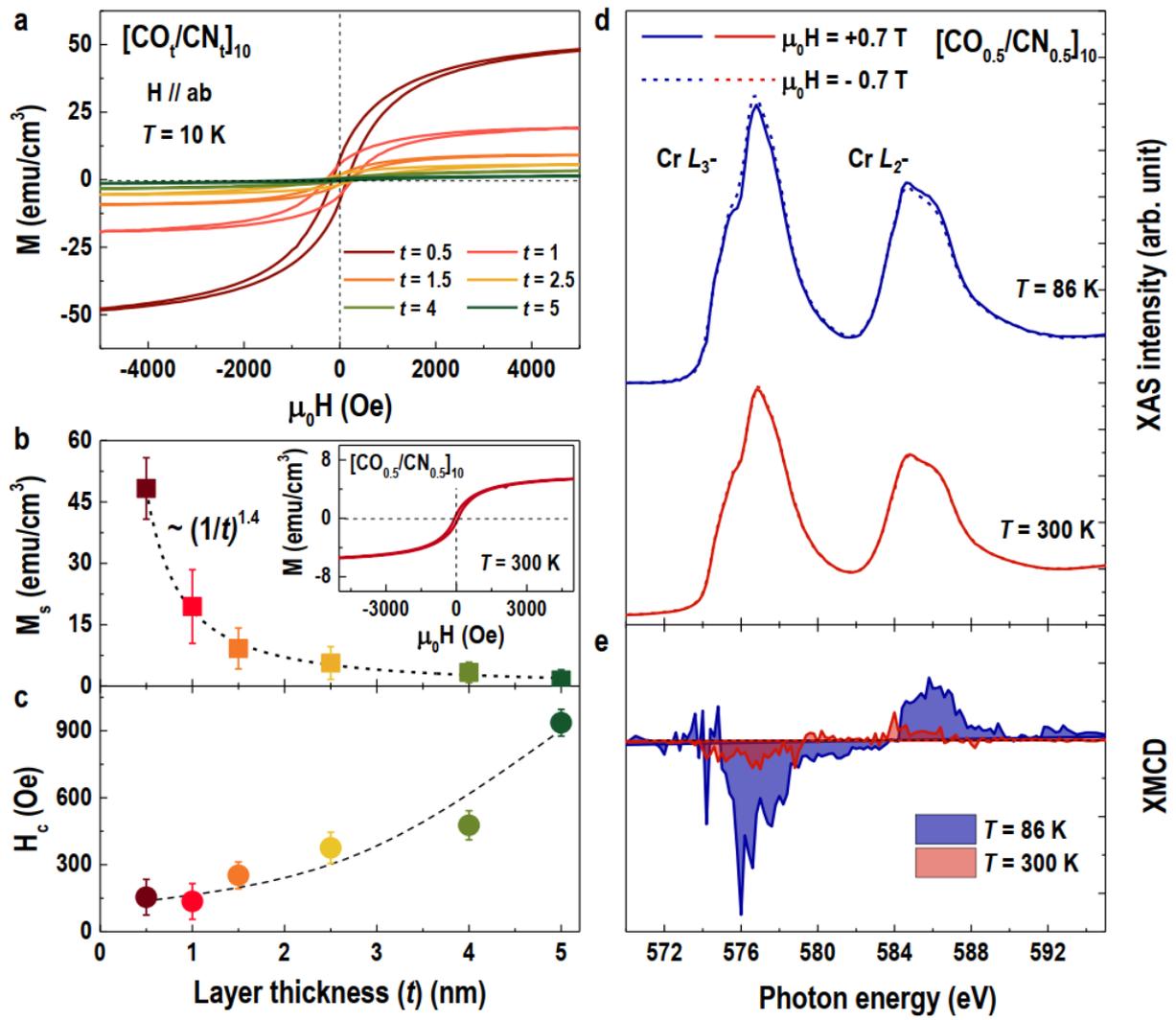

**Figure 3.**

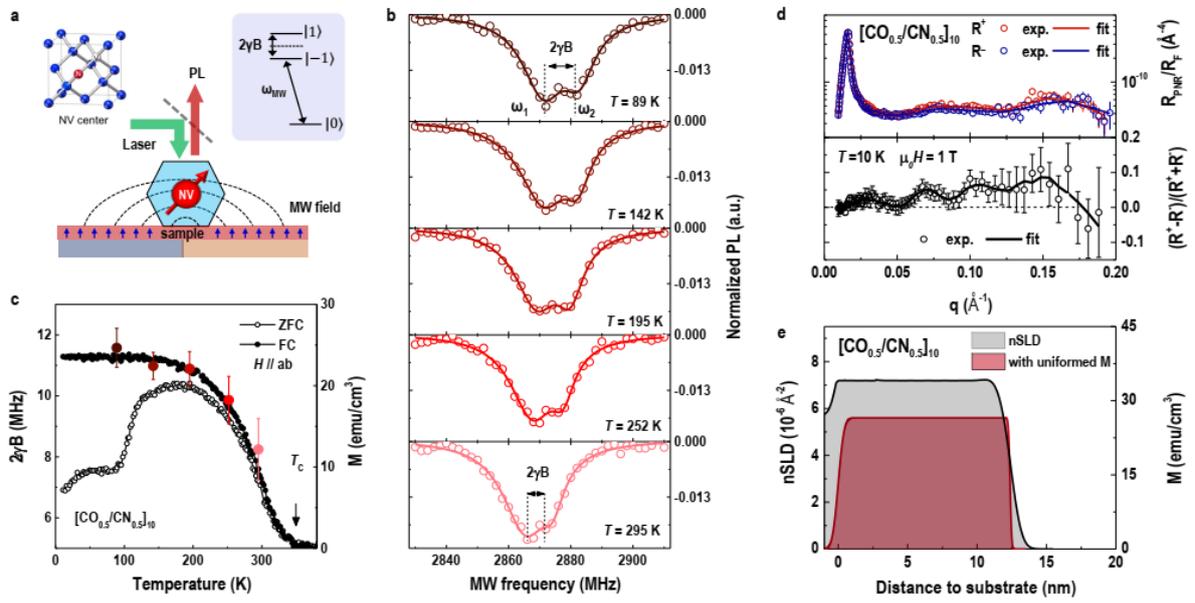

**Figure 4.**

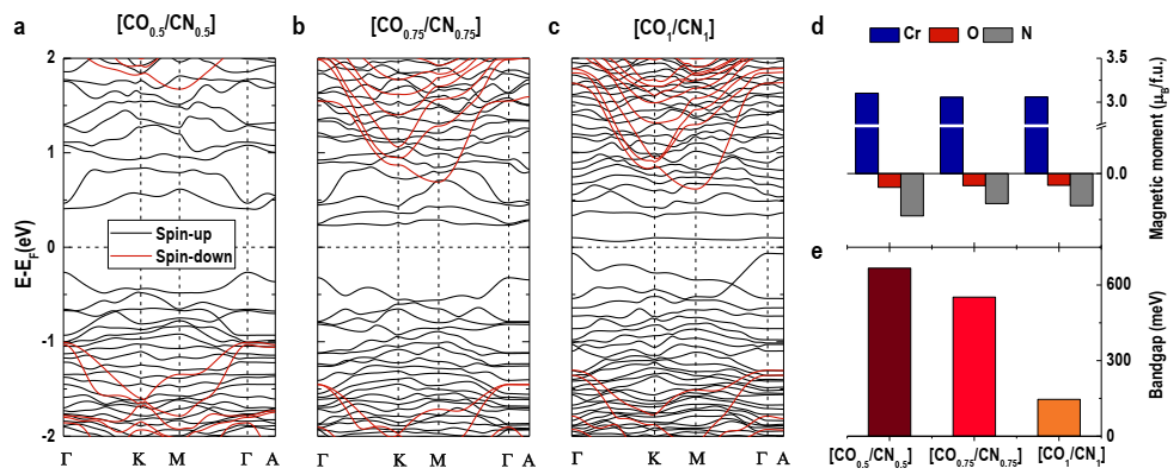